\documentclass{article}

\PassOptionsToPackage{numbers, compress}{natbib}

\usepackage[final]{neurips_2020}
\usepackage[utf8]{inputenc} 
\usepackage[T1]{fontenc}    
\usepackage{hyperref}       
\usepackage{url}            
\usepackage{booktabs}       
\usepackage{amsfonts}       
\usepackage{nicefrac}       
\usepackage{microtype}      
\usepackage{graphicx}
\usepackage{tabularx}
\usepackage{multirow}
\usepackage{color}
\title{Deep Sequence Learning for Accurate Gestational Age Estimation from a $\$$25 Doppler Device}

\author{%
  Nasim Katebi\\
  Department of Biomedical Informatics\\
  Emory University School of Medicine\\
  Atlanta, GA, USA \\
  \texttt{nkatebi@emory.edu} \\
  \And
  Reza Sameni \\
Department of Biomedical Informatics\\
  Emory University School of Medicine\\
  Atlanta, GA, USA \\
  \texttt{rsameni@dbmi.emory.edu } \\
   \AND
   Gari D. Clifford \\
   Department of Biomedical Informatics\\
   Emory University School of Medicine\\
   Department of Biomedical Engineering\\
   Georgia Institute of Technology\\
   Atlanta, GA, USA\\
  \texttt{gari@gtech.edu} \\
}

\begin{document}

\maketitle

\begin{abstract}
Assessing fetal development is usually carried out by techniques such as ultrasound imaging, which is generally unavailable in rural areas due to the high cost, maintenance, skills and training needed to operate the devices effectively. In this work, we propose a low-cost one-dimensional Doppler-based method for estimating gestational age (GA). Doppler time series were collected from 401 pregnancies between 5 and 9 months GA using a smartphone. The proposed model for GA estimation is based on sequence learning by forming a temporally dependent model using a convolutional long-short-term memory network. Time-frequency features are extracted from Doppler signals and regularized before feeding to the network. The overall mean absolute GA error with respect to the last menstrual period was found to be 0.71 month, which outperforms all previous works.
\end{abstract}

\section{Introduction}
 Low-and middle-income countries (LMICs) account for approximately 98\% of all reported perinatal deaths worldwide, mainly due to gestational developmental issues, specifically intrauterine growth restriction (IUGR) \cite{zupan2005perinatal,lee2013national,cuadros2020review}. Rising costs of healthcare and inadequate access to prenatal medical services exacerbate this issue in LMICs as well as low-income regions in developed countries, such as the southeast US. Most of these deaths can be avoided by improving health monitoring before, during and after childbirth. Therefore, developing AI-enabled edge-computing devices that are intuitive to use, even for low-literacy populations helps to enhance healthcare for disadvantaged populations. 
 
Gestational age estimation provides essential information such as preterm birth management, delivery scheduling and growth restriction \cite{alexander1995discordance}. Fetal cardiac assessment is a tool recommended by obstetrical societies for monitoring fetal health during pregnancy \cite{liston2007fetal}. The functional assessment of the fetal heart conveys important information regarding the hemodynamic status and cardiovascular adaptation of a fetus in the face of several perinatal complications. Fetal heart rate is influenced by the autonomic nervous system (ANS), which matures during pregnancy. In particular, fetal heart rate variability evolves over the course of pregnancy reflecting the maturity of the ANS, and thus an indirect indicator of the fetal gestational age \cite{wakai2004assessment}. Previous studies have shown that fetal heart rate variability metrics can be used as discriminative features for fetal development assessment \cite{valderrama2020proxy,hoyer2013fetal,tetschke2016assessment,marzbanrad2016estimating,marzbanrad2017assessment}. One non-invasive method for capturing fetal cardiac activity is one-dimensional Doppler ultrasound (1D-DUS), which is a low-cost and simple method for fetal heart rate monitoring \cite{marzbanrad2018cardiotocography}. The Doppler transducer can be easily adapted to connect to mobile devices such as smartphones, for recording and processing, motivating their use in mobile-health (mhealth) systems for risk screening in low-resource environments \cite{stroux2016mhealth}. The Doppler transducer transmits and receives ultrasound waves, which reflect fetal cardiac activity. Using 1D-DUS signal, blood flow, cardiac wall, and valve motions can be captured and are differentiable based on their different velocities. Although 1D-DUS provides useful information regarding cardiac functionality, its variable morphology makes the processing and modeling of this signal challenging.

In this paper, we propose a systematic approach for assessing fetal development by discovering the relation between fetal 1D-DUS signal and gestational age. The proposed approach is based on a set of effective time-frequency domain features of the 1D-DUS and a convolutional long-short-term memory (CLSTM) network \cite{xingjian2015convolutional}, which is a robust and powerful method for extracting features from sequential data. This approach is used to model time dependencies in fetal 1D-DUS and to capture the variability of the cardiac activity, eventually leading to the estimation of the fetal gestational age. In the sequel, we start by formulating a mathematical model for the problem of fetal gestational age estimation from 1D-DUS signals.

\begin{figure}
  \centering
  	\includegraphics [trim={1.5cm 10cm 5cm 0cm},width=1.0\textwidth, clip]{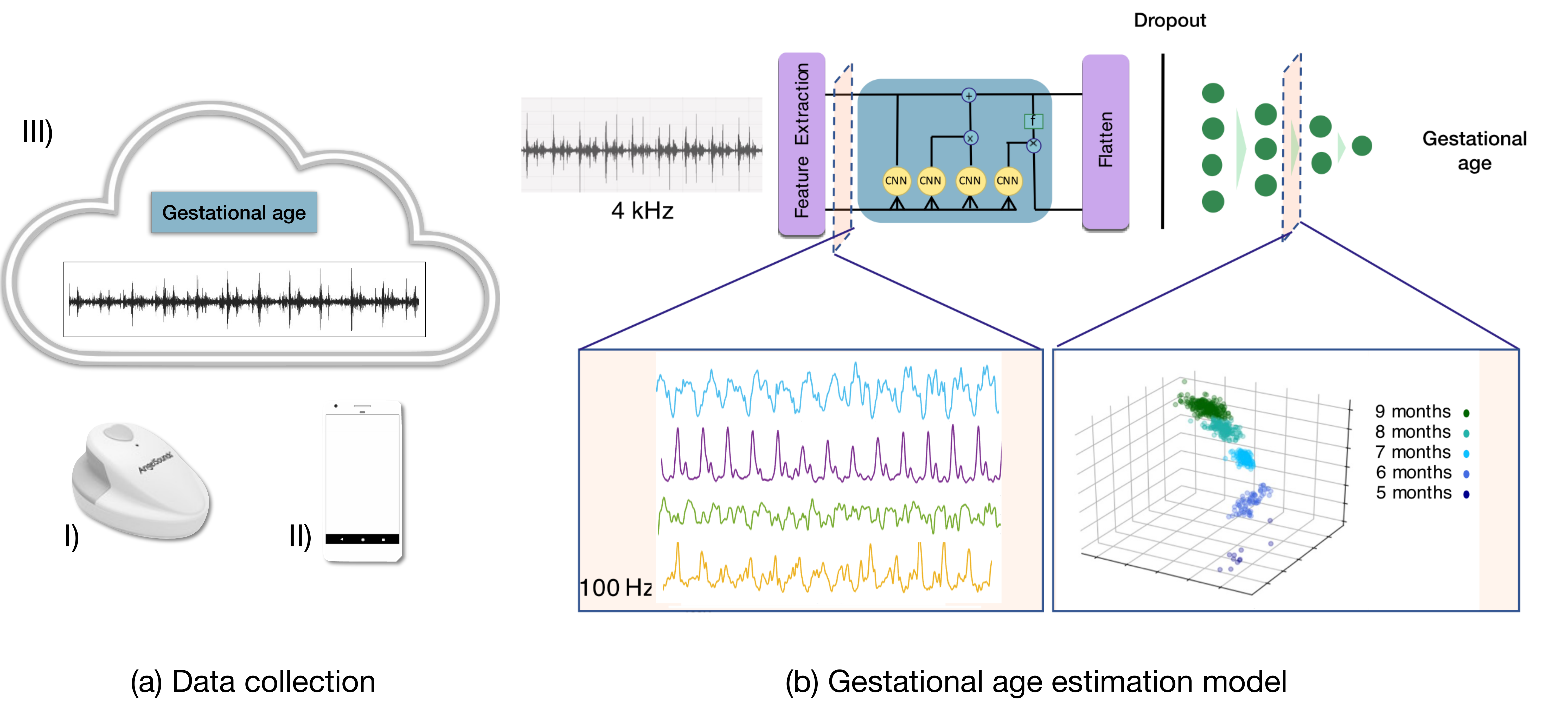}
  \caption{(a) Data collection. I) The raw 1D ultrasound is captured using Doppler transducer and II) The 1D-DUS and gestational age are recorded on the phone using the developed mobile app. III) The data are then uploaded to the cloud for backup and further processing. (b) An overview of the proposed process for training CLSTM network for fetal monitoring from abdominal Doppler acquired during routine fetal monitoring. The features from 1D-DUS are calculated and fed to the CLSTM network. The output is then flattened and mapped to the target label.}
\label{fig:GA_model}  
\end{figure}

\section{Gestational age estimation model}


Let $x(n;t)$ denote the time series of a 1D-DUS signal with discrete time index $n$, acquired during a clinical visit of a pregnant woman on date $t$. For simplicity, the date and gestational age are represented in units of weeks. The ``true gestational age'' at date $t$ is denoted $a(t) = t - c$, where $c$ is the \textit{date of conception}, while the presumed or reported gestational age is $\tilde{a}(t) = t - \tilde{c}$, where $\tilde{c}$ is the \textit{anticipated conception date}. 
Therefore, the presumed (anticipated) and true gestational ages can be related as follows:
\begin{equation}
    \tilde{a}(t) = a(t) + \eta
\end{equation}
where $\eta = c - \tilde{c}$ is the gestational age presumption error, which without additional priors (such as 2D-Doppler) remains an unknown stochastic constant over pregnancy. The error $\eta$ accounts for lack of knowledge of the last menstrual period and uncertainties in the exact ovulation, intercourse and conception dates. We further denote the $p$-dimensional feature vector extracted from the 1D-DUS by $\mathbf{f}(x(n;t)) \in \mathbb{R}^{p}$. The objective is to design a deep network that estimated the true gestational age from the feature vector extracted from a single or a set of 1D-DUS acquired during pregnancy, i.e.,
\begin{equation}
    \hat{a}(t) = \displaystyle G\left[\tilde{a}(t), \{\mathbf{f}(x(n;t_k))\}_{k = 1}^L\right]
\end{equation}
where $\hat{a}(t)$ is an estimate of the true gestational age, $t_k$ ($k = 1, \ldots, L$) denote the $L$ dates that 1D-DUS is acquired from the pregnant woman, and $G(\cdot)$ denotes the feature-vector to gestational age transform that is learned by the neural network, as shown in Fig.~\ref{fig:GA_model}-(b). In this scheme, the presumed gestational age $\tilde{a}(t)$ is used for model training.

\section{Dataset}
A hand-held 1D-DUS device, the AngelSounds Fetal 1D-DUS JPD-100s (Jumper Medical Co., Ltd., Shenzhen, China), with an ultrasound transmission frequency of 3.3~MHz, and costing \$25, was used to capture audio data from 401 pregnant women (493 visits and 693 recordings) at 5 to 9 months of gestation. 
The data, collected as part of a randomized control trial conducted in rural highland Guatemala \cite{martinez2018mhealth,valderrama2018improving}, 
include 15, 77, 162, 186, 253 recordings corresponding to gestational ages of 5, 6, 7, 8 and 9 months, respectively.
The 1D-DUS signals were recorded by traditional birth attendants, who were trained to use the hand-held 1D-DUS device and were an accompanying mobile application. Immediately before recording the 1D-DUS signals, the traditional birth attendants entered the estimated gestational age  into the app in months, based on the last menstrual period (LMP). Data were captured using a bespoke Android client at 44.1~kHz, using a low-cost smartphone (Samsung S3 mini) and stored as uncompressed WAV files at 7056/s bits) \cite{stroux2016mhealth}. The first five minutes of each recording were used. Figure \ref{fig:GA_model}-(a) illustrates the data sources and devices used in this research.

\section{Data analysis and feature extraction}

\subsection{Preprocessing}
Given the nature of the physiological time-series, 1D-DUS signals are corrupted with internal and external interference such as respiration, movement, and environmental noise. In this work, a second-order band-pass Butterworth filter was used to reduce the noise. By observing the frequency components of the 1D-DUS signals, the cut-off frequencies were set to 25 and 600~Hz, corresponding to cardiac oscillations. 
\subsection{Time-frequency (TF) features for DUS components}
\label{sec:TFDUSFeature}

For a real-valued discrete-time signal $x_n$, where $n$ is the time instant, we define a windowed version of the signal $s_n = w_n x_n$, 
where $w_m$ ($m=0, \ldots, N-1$) is a window for improving the spectral features and minimizing the windowing effects. A Hamming window of length 100~ms (400 samples at a sampling rate of 4~kHz) is used for the later presented results. The discrete-time Fourier transform (DTFT) of a window of $N$ samples of $s_n$ is:
\begin{equation}
\displaystyle S_n(\omega) \stackrel{\Delta}{=} \displaystyle \sum_{m = 0}^{N-1}  \displaystyle s_{n+m-N+1} e^{-j\omega m}
\label{eq:DTFT}
\end{equation}
According to the Parseval's theorem, the energy of each window of the signal is:
\begin{equation}
E_n \stackrel{\Delta}{=} \sum_{m = 0}^{N-1} |s_{n-m}|^2 = \frac{1}{\pi} \int_{0}^{\pi} |S_n(\omega)|^2 d\omega
\label{eq:energy}
\end{equation}
We define the \textit{instantaneous frequency}, or the \textit{first spectral moment} of $s_n$, as follows:
\begin{equation}
\omega_n \stackrel{\Delta}{=} \frac{1}{E_n} \int_{0}^{\pi} \omega |S_n(\omega)|^2 d\omega
\label{eq:freq}
\end{equation}
which for frequency-domain unimodal signals is the frequency (in radians) around which the signal energy is localized at time instant $n$. It is therefore a measure of the signal's center frequency at time $n$. In a similar manner, the \textit{instantaneous bandwidth}, or the \textit{spectral centralized second moment}, of $s_n$ is defined:
\begin{equation}
\Delta\omega_n^2 \stackrel{\Delta}{=} \frac{1}{E_n} \int_{0}^{\pi} (\omega - \omega_n)^2 |S_n(\omega)|^2 d\omega
\label{eq:bandwidth}
\end{equation}
which is a measure of the instantaneous energy spread around the instantaneous frequency.

Finally, we define the instantaneous oscillation quality-factor (\textit{Q-factor}): $Q_n \stackrel{\Delta}{=} \omega_n/\Delta\omega_n$, 
as a measure of oscillation quality, which is a notion commonly used in electronic circuitry for evaluating the quality of oscillation independent of the frequency. Accordingly, when $x_n$ becomes closer to a single-tone component, the Q-factor increases. Note that for digital implementations, the DTFT is replaced by the Discrete Fourier Transform (DFT), with appropriate dimension corrections. 
These features form the overall feature vector $\mathbf{f}_n \stackrel{\Delta}{=} (\sqrt{E_n}, \omega_n, \Delta\omega_n^2, Q_n)$, which is fed to the sequence modeling part of the model. Due to the signal windowing, the extracted feature vector has slow variations over time and was therefore resampled from 4~kHz to 100~Hz, to reduce the processing load.

\subsection{Sequence modeling}
Given that we have a sequential feature vector, the use of a recurrent neural network is a natural choice to keep track of the variability and temporal structure of the signal. Long-short-term memory (LSTM) \cite{hochreiter1997long} is one such recurrent neural network, which has been used in various studies for the general purpose of sequence modeling. A Convolutional LSTM (CLSTM) network developed by Shi et al. \cite{xingjian2015convolutional} is a combination of LSTM and convolutional neural network to capture spatio-temporal features. Accordingly, the input-to-state and state-to-state transitions in LSTM are changed from full connections to convolution structure. By stacking multiple CLSTM layers, one can form a spatio-temporal sequence modeling network to uncover the variability in fetal cardiac activity. By changing the kernel size, CLSTM is able to capture the different DUS components with different velocities, corresponding to the different fetal-maternal body tissues that move within the DUS transceiver frequency range. We can consider the states as the hidden representation of the cardiac valve opening and closing. These sub-organ motions are captured by setting the appropriate kernel size. The utilized CLSTM architecture comprised of two successive layers with kernel sizes $(1,20)$ and $(1,10)$, respectively. Each layer is followed by batch normalization. The output is then flattened and mapped to the gestational age label through 4 fully connected layers with sizes 128, 32, 3 and 1. In order to reduce the likelihood of over-fitting we used an L2-regularizer with regularization parameter $\lambda=0.01$. A dropout technique was also used before the dense layer and the probability of training a given node in a layer was set to 0.3.  

\section{Results} Stratified five-fold cross-validation is used across patients to assess the performance of gestational age estimation. The model was trained end-to-end for a total of 300 epochs using a mean absolute error (MAE) loss function. The batch size was fixed to 32 patients and generated using a balanced batch generator with random oversampling (with replacement) of the less frequent label (5 months). The 50 trial cross-validation was performed and the median, the lower and upper 95\% confidence interval (LCI and UCI ) of MAE values were determined (Table~\ref{table:GA_estimations_trials}). Our proposed model outperforms the previous studies on gestational age estimation, which were based on 1D-DUS signals and maternal blood pressure and heart rate \cite{valderrama2020proxy} and 1D-DUS signals with simultaneously recorded fetal electrocardiogram (ECG) \cite{marzbanrad2016estimating,marzbanrad2017assessment}. It should be noted that although using simultaneously recorded ECG or maternal blood pressure can improve the performance of the gestational age estimation \cite{valderrama2020proxy,rebelo2015blood}, for the application of interest, requiring an additional device for blood pressure or ECG recordings significantly complicates the use and raises the cost of the smartphone-mediated perinatal screening system.
\begin{table}[!ht]
\small{
\caption{ Mean absolute errors of the 50-trial five-fold cross validation for the CLSTM in months. Error is reported as lower, median and upper 95\% confidence interval (LCI, median, UCI) for GAs of 5-9 months, together with the average over all months tested (All).}
\label{table:GA_estimations_trials}
\centering
\begin{tabular}{ccccccc} 
\toprule
     \multirow{2}{*}{}  & \multicolumn{5}{c}{Gestational Age (months since reported LMP)} \\
    & 5&  6 & 7  & 8 & 9 & All \\ 
 \midrule    
Error	 &  (1.9, 1.98, 2.1) &(0.7, 0.72, 0.8) &	(0.4, 0.45, 0.5) &	(0.4, 0.48, 0.4) &	(0.9, 0.98, 1.1) &	 	0.71\\
	 \bottomrule
\end{tabular}
}
\end{table}

\section{Conclusion}
This work represents the first attempt to estimate gestational age from only Doppler signals, and outperforms previous attempts based on multiple signals (Doppler plus electrocardiogram \cite{marzbanrad2016estimating} or Doppler plus blood pressure \cite{valderrama2020proxy}). Since the error is close to the quantization of the labels, future improvements will require more accurate GA labels, collected using Doppler imaging in the first trimester.

\section{Acknowledgements}
GC acknowledges the support of the National Institutes of Health, the Fogarty International Center and the Eunice Kennedy Shriver National Institute of Child Health and Human Development, grant number 1R21HD084114-01 (Mobile Health Intervention to Improve Perinatal Continuum of Care in Guatemala). GC has financial interest in Alivecor Inc, and receives unrestricted funding from the company. GC also is the CTO of Mindchild Medical and has ownership interests in Mindchild Medical. RS has equity interests in Mindchild Medical.

\bibliographystyle{ieeetr}
\bibliography{ref}






\end{document}